\documentclass{PoS}
\usepackage[centertags]{amsmath}
\usepackage{amsfonts} \usepackage{amssymb} \usepackage{amsthm}
\usepackage{graphicx}
\usepackage{psfrag}

\def\one{{\rm 1\kern -.9mm l}}                             %

\def\beq{\begin{equation}}
\def\eeq{\end{equation}}
\def\beq{\begin{equation}}
\def\eeq{\end{equation}}
\def\beqa{\begin{eqnarray}}
\def\eeqa{\end{eqnarray}}
\newcommand{\eqa}{\begin{eqnarray}}
\newcommand{\ena}{\end{eqnarray}}
\newcommand{\eq}[1]{eq. (\ref{#1})}

\def\ee{\mathrm{e}}

\title{Effective string theory description of the interface free energy}

\ShortTitle{Effective string theory description of the interface free energy}

\author{Marco Bill\'o,~~~\speaker{Michele Caselle}, Livia Ferro\\
Dipartimento di Fisica Teorica, Universit\`a di Torino\\
and Istituto Nazionale di Fisica Nucleare - sezione di Torino,\\
Via P. Giuria 1, I-10125 Torino, Italy\\
        E-mail: \email{billo,caselle,ferro@to.infn.it}}

\author{Martin Hasenbusch\\Dipartimento di Fisica dell'Universit\`a di Pisa
 and I.N.F.N.,\\
 Largo Bruno Pontecorvo 3, 
                  I-56127 Pisa,
                        Italy  \\
        E-mail: \email{Martin.Hasenbusch@df.unipi.it}}

\author{Marco Panero\\Institute for Theoretical Physics,
University of Regensburg,\\
93040 -- Regensburg,
                              Germany \\
        E-mail: \email{marco.panero@physik.uni-regensburg}}

\abstract{We compare the predictions of the Nambu-Goto effective string model with a set of high precision 
Monte Carlo results for interfaces with periodic boundary conditions in the 3D Ising model. We compute the 
free energy in the covariant gauge exactly, up to the inclusion of the Liouville mode. The perturbative 
expansion of this result agrees both with the result evaluated several years ago by Dietz and Filk 
in the physical gauge and with a recent calculation with the 
Polchinski-Strominger action. We also derive the effective string spectrum which, because of the different 
boundary conditions, is very different from the well known one of Arvis. Taking into proper account the 
effective string corrections and exploiting some technical improvements in the simulations we obtain  
precise estimate of the amplitude ratios 
$\frac{T_c}{\sqrt{\sigma}}$,
$\frac{m_{0++}}{\sqrt{\sigma}}$ and
$\sigma \xi_{2nd}^2 $. 
We also discuss the behaviour of the effective string 
free energy in the dimensional reduction limit (i.e., near the deconfinement transition of the dual 3d 
gauge Ising model) and its relationship with the 2d Ising model interfaces}

\FullConference{The XXV International Symposium on Lattice Field Theory\\
		 July 30 - August 4 2007\\
		 Regensburg, Germany}

\begin{document}

\section{Introduction}
The properties of interfaces in three-dimensional statistical systems have been a 
long-standing subject of research. In particular the interest of people 
working in the subject has been attracted by the so called ``fluid'' interfaces
whose dynamics is dominated by massless excitations
(for a review see for instance \cite{p92}). For 
this class of interfaces, thanks to the presence of long range massless modes,
microscopic details such as the  lattice structure of the spin model or the
chemical composition of the components of the binary mixture 
become irrelevant and the physics  can be rather accurately 
described by field theoretic methods. 

A simple realization of these fluid interfaces is represented by 3d spin models.
These models in the low temperature phase in which their global symmetry is
broken admit different vacua which, for a suitable choice of the boundary
conditions, can  occupy macroscopic regions and are 
separated by domain walls which behave as interfaces. For
temperatures  between the roughening and the critical one, 
interfaces are dominated by long wavelength fluctuations (i.e.\ they exactly  behave as
{\em  fluid} interfaces).

An effective model widely used to describe a
rough interface is the {\em  capillary wave model} (CWM) \cite{BLS,rw82}. 
Actually this model (which was proposed well before the Nambu-Goto papers~\cite{Goto:1971ce,Nambu:1974zg})
exactly coincides with the Nambu-Goto one, since 
it assumes an effective Hamiltonian
proportional to the variation of the surface's area with respect to the
classical solution. 

Among the various realizations of fluid 
interfaces, a prominent role has been played in these last years by the 
3d Ising model, for several reasons. The universality class of the Ising model
includes many physical systems, ranging from binary 
mixtures to amphiphilic membranes.  It also shares the universality class with the $\phi^4$ theory;
this allows a QFT approach to the description of 
the interface physics.
Last but not least, the Ising model, due to 
its intrinsic simplicity, allows fast and high statistics Monte Carlo
simulations, so that very precise 
comparisons can be made between theoretical predictions and numerical results.

Another important reason of interest of interfaces in the 3d Ising model is that, thanks to the duality relation
with the 3d gauge Ising model, they are deeply related with  Wilson loop observables. 
This is the
ultimate reason for the validity of the Nambu-Goto effective string model for interfaces. Indeed 
this mapping allows to
create a dictionary between Lattice Gauge Theory (LGT) observables and interface ones and makes the results
obtained in the context of interface physics immediately relevant also for LGT studies. 
In this respect, it is important to stress
that by working directly in the 3d Ising model one has a larger freedom in the choice of the boundary conditions.
In this work we shall exploit this freedom and address interfaces with periodic boundary conditions. 
This choice is particularly useful if one wants to address 
subtle finite size corrections (as one has to do in order to test the Nambu-Goto effective string picture)
since the periodic b.c. allow to eliminate any source of
boundary corrections in the results, thus leading to a very clean and artefact-free comparison between theory and
simulations.

The aim of this contribution is to summarize a set of results recently obtained in this context 
by our group~\cite{bcf06,chp06,chp07,bcf07}. In particular we shall discuss the following items:
\begin{description}
\item{1]} Explicit evaluation of the Nambu-Goto partition function (neglecting the anomaly) for
interfaces~\cite{bcf06}.
 
\item{2]} A collection of numerical results from a set of high precision Montecarlo
simulations~\cite{chp06,chp07}

\item{3]} The 2d limit of  3d interfaces~\cite{bcf07}.

\end{description}

\section{The partition function of interfaces from the Nambu-Goto effective string}

The Nambu-Goto bosonic string is defined by an action proportional, via the string tension $\sigma$, to the
induced area of a surface embedded in a $d$-dimensional target space:
\begin{equation}
\label{ngaction}
S = \sigma \int d^2\xi \sqrt{\det g}~,
\hskip 0.6cm g_{\alpha\beta} = \frac{\partial X^i}{\partial\xi^\alpha}\frac{\partial X^j}{\partial \xi^\beta} G_{ij}~,  
\end{equation}
where the proper coordinates $\xi^\alpha$ parametrize the world sheet while
$X^i(\xi)$ ($M=0,\ldots,d$) describes the target space position of a point specified by $\xi$.
We shall assume in the following that the target space metric $G_{ij}$ is the flat one and neglect it. 

The standard approach to evaluate the partition function is to fix the re-parametrizations of the action
by fixing the  so-called ``physical'' gauge in which the proper coordinates are identified with two 
of the target space coordinates, say $X^0$ and $X^1$. The quantum version of the NG theory 
can then be defined through the functional integration over the $d-2$ transverse 
d.o.f. $\vec X(X^0,X^1)$ of the gauge-fixed action. 
The partition function for a surface $\Sigma$ with prescribed boundary conditions
is given by
\begin{equation}
\label{ngpart}
\begin{aligned}
Z_{\Sigma} & = \int_{(\partial\Sigma)}\!\!\!\! DX^i\, \exp\left\{-\sigma \int_\Sigma dX^0 dX^1 \left(1 + (\partial_0\vec X)^2
 + (\partial_1\vec X)^2 + (\partial_0 \vec X \wedge \partial_1 \vec X)^2\right)^{\frac 12}\right\}
\\
& = \int_{(\partial\Sigma)}\!\!\!\! DX^i\,  \exp\left\{-\sigma \int_\Sigma dX^0 dX^1 \left[
1 +\frac 12 (\partial_0\vec X)^2 + \frac 12 (\partial_1\vec X)^2 + \mbox{interactions} \right]\right\}~.
\end{aligned}
\end{equation}

Expanding the square root as in the second line above, the classical area law $\exp(-\sigma\mathcal{A})$ 
(where $\mathcal{A}$ is the area of the minimal surface $\Sigma$) is singled out. It multiplies
the quantum fluctuations of the 
fields $\vec X$, which have a series of higher order (derivative) interactions.
The functional integration can be performed perturbatively (since we are studying an effective model we can
neglect the fact that the theory is actually non-renormalizable), the loop expansion parameter being 
$1/(\sigma\mathcal{A})$, and it 
depends on the boundary conditions imposed on the fields $\vec X$, i.e., on the topology of the
 boundary $\partial\Sigma$ and hence of $\Sigma$. The cases in which $\Sigma$ is a disk, a cylinder 
 or a torus are
the ones relevant for an effective string description of, respectively, Wilson loops, 
Polyakov loop correlators and interfaces in a compact target space. 
The computation was carried out up to two loops in \cite{df83}.

The main problem of this approach is that the physical gauge is anomalous in $d\not=26$ dimensions.
However it can be shown~\cite{o85} 
that the anomaly decreases as $R^{-3}$, where 
$R$ is the typical scale of the boundary of $\Sigma$ and thus it could be possible that it does
not affect the first two terms in the loop expansion.

In~\cite{bcf06} we proposed an alternative treatment of the NG model which
takes advantage of the first order formulation, in which the action is simply 
\beq
\label{sac}
S = \sigma\int d\xi^0 \int_0^{2\pi} d\xi^1\,
h^{\alpha\beta}\partial_\alpha X^i \partial_\beta X^i~,
\eeq
where $h_{\alpha\beta}$ is an independent world-sheet metric, $\xi^1\in [0,2\pi]$ parametrizes the 
spatial extension of the string and $\xi^0$ its proper time evolution. 
Integrating out $h$, we retrieve the NG action \eq{ngaction}.
For each topology of the world-sheet, classified simply by its genus $g$,
we can instead use re-parametrization and  Weyl invariance to put the metric in a reference form 
$\ee^\phi \hat h_{\alpha\beta}$ (conformal gauge fixing). For instance, on the sphere,
 i.e. at genus $g=0$, we can choose $\hat h_{\alpha\beta} = \eta_{\alpha\beta}$, 
 while on the torus, at genus $g=1$, $\hat h_{\alpha\beta}$ is constant, 
 but still depends on a single complex parameter $\tau$, 
 the modulus of the torus.
The scale factor $\ee^\phi$ decouples at the classical
level and
the action takes then the form
\begin{equation}
\label{sac2}
S = \sigma\int d\xi^0 \int_0^{2\pi} d\xi^1\,
\hat h^{\alpha\beta}\partial_\alpha X^i \partial_\beta X^i
 + S_{\mathrm{gh.}}~,
\end{equation}
where $S_{\mathrm{gh.}}$ in \eq{sac} is the action for the ghost
and anti-ghost fields (traditionally called $c$ and $b$) that arise from the
Jacobian to fix the conformal gauge; we do not really need here its explicit
expression, the only important thing to know is that they correspond
to a CFT of central charge $c_{\mathrm{gh.}} = -26$.
The fields $X^i(\tau,\sigma)$, with $i=1,\ldots,d$, describe the embedding of
the string world-sheet in the target space and form the simple, well-known two-dimensional CFT of $d$
free bosons. This allows to evaluate exactly the sum over all possible surfaces for any value of $d$.
In the particular case of
the toroidal geometry (interfaces) in which we are interested
one finds (see~\cite{bcf06} for
the details):
\begin{equation}
\label{boskkp}
\mathcal{I}^{(d)} = 2 \left(\frac{\sigma}{2\pi}\right)^ {\frac{d-2}{2}}\, V_T \, \sqrt{\sigma\mathcal{A}u}
\sum_{k,k'=0}^\infty  c_k c_{k'} 
\left(\frac{E}{u}\right)^{\frac{d-1}{2}}\, K_{\frac{d-1}{2}}\left(\sigma\mathcal{A}E\right)~,
\end{equation}
where $\mathcal{A} \equiv L_1 L_2$
 denotes the minimal area of the interface, $u \equiv \frac{L_2}{L_1}$ is the ratio between the two lattice sizes
of the plane in which the interface lies ($L_i$ denotes the lattice sizes in the $i$th direction),
 $V_T = \prod_{i=3}^d L_i$ is the transverse
volume, $c_k$ is the $k$th coefficient in the expansion of the Dedekind function:
 \begin{equation}
\label{bos14}
\left[\eta(q)\right]^{2-d} = \sum_{k=0}^{\infty}c_k q^{k-\frac{d-2}{24}}~,
\end{equation}
$K_{\frac{d-1}{2}}$ is the modified Bessel function of order $\frac{d-1}{2}$ and

\begin{equation}
\label{bos19}
E  =
 \sqrt{1 + \frac{4\pi\, u}{\sigma \mathcal{A}}(k+k'-\frac{d-2}{12}) + 
\frac{4\pi^{2} \, u^2\, n_{1}^{2}}{(\sigma\mathcal{A})^2}}~.
\end{equation}
denotes the spectrum of the NG string for this particular type of geometry.

It is important to stress that also in this case for $d\not=26$ we must expect deviations from this result.
In fact the decoupling of the scale factor $\ee^\phi$ persists at the quantum level only if the anomaly
parametrized by the total central charge $c= d - 26$ vanishes;
Weyl invariance is otherwise broken and the mode $\phi$
has to be thought of as a field with a Liouville-type action.

Notwithstanding this there are a few non-trivial results of this analysis which are not affected by the anomaly
and deserve further attention:
\begin{itemize}
\item
Eq.(\ref{boskkp}) gives an explicit prediction for the functional dependence on $\sigma,\mathcal{A},u$ and $V_T$
of the prefactor in the
interface free energy which (as we shall see in the next section) nicely agrees with the results of the MC
simulations.
\item
The spectrum reported in eq.(\ref{bos19}) is different from the one obtained by Arvis in~\cite{arvis}
for the Wilson loop geometry and agrees with an independent calculation mentioned in~\cite{Kuti:2005xg}.

\item 
Eq.(\ref{boskkp}) in the $d=3$ case allows a smooth dimensional reduction limit and suggest an expression for the
2d interface which agrees with an exact 2d calculation.
\item
Expanding eq.(\ref{boskkp}) in powers of $1/(\sigma\mathcal{A})$ we reobtain at the 
first two orders the same
contributions obtained in~\cite{df83} with the physical gauge.
This is probably the most interesting result since it supports the idea that not only the first
term in the expansion (the so called "L\"uscher term"~\cite{lsw80}) but also the next to leading
order could be anomaly free and universal. This conjecture is further supported by the recent
observation~\cite{d04,hdm06} that the same term also appears in the expansion of the 
Polchinski-Strominger action~\cite{ps91} which is essentially a non-local, but anomaly free,
version of the N-G action. 

\end{itemize}
 
\begin{figure}
\hskip -0.5cm
\includegraphics[height=10cm]{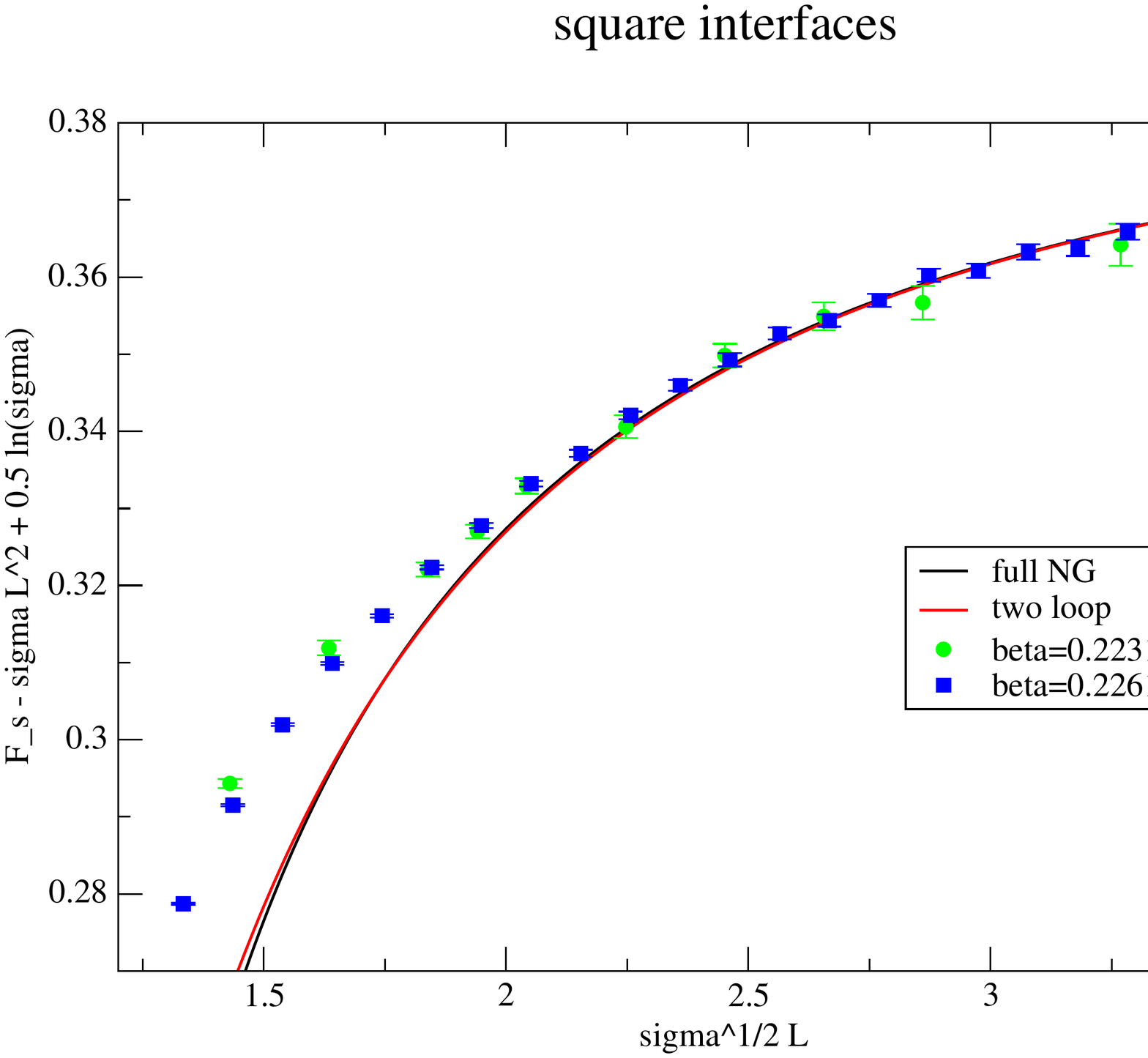}
\includegraphics[height=10cm]{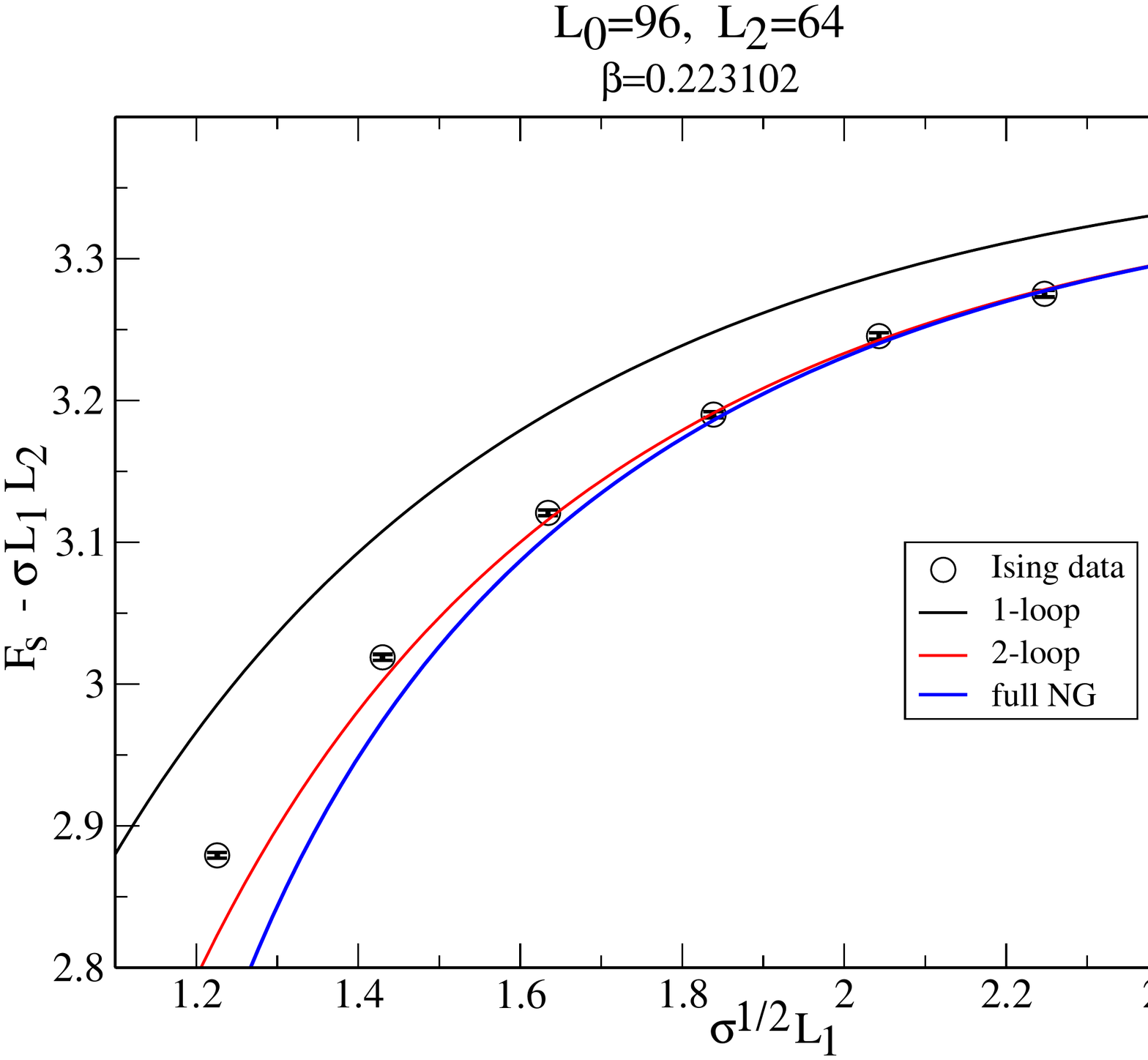}
\vskip0.5cm
\caption{{\bf Fig.1a}: Comparison of the 2-loop prediction, the full Nambu-Goto result with the data for square 
interfaces at $\beta=0.223102$ and $\beta=0.226102$. 
{\bf Fig.1b}: $F_{\mbox{\tiny{s}}}^{(2)}- \sigma L_1 L_2$ as a function of 
$\sqrt{\sigma} L_1$ for $\beta=0.223102$, $L_3=96$ 
and $L_2=64$. Note that in the case of the Monte Carlo results the statistical error 
is smaller than the symbol (circle). The 1-loop, 2-loop and full NG predictions are 
given as solid black, red and blue lines, respectively.}
\end{figure}

\section{Monte Carlo simulations of interfaces in the three-dimensional Ising model}
In order to test the above results it is mandatory to have very precise estimates of the interface free energy
for different values of $L_1$ and $L_2$. In~\cite{chp06} and \cite{chp07} we obtained these estimates using two 
different approaches. In~\cite{chp06} we used the "boundary flip'' algorithm~\cite{Hasenbusch:1992eu} which is
particularly suited for interfaces of small or intermediate size while in~\cite{chp07} we obtained the interface
free energy by a direct integration of the differences $E_a-E_p$ over $\beta$
(where $E_a$ denotes the internal energy for
antiperiodic boundary conditions, and hence in presence of an odd number of interfaces, while $E_p$ denotes the
internal energy with periodic b.c.).  This second method is much more
cumbersome from a computational
point of view, but it is mandatory if one is interested in
large values of the interface free energy. 
We refer to the original papers for all the details on the algorithms and on the simulation settings
and only list here the main results.
\begin{itemize}

\item
As it was already observed in the case of Wilson loop~\cite{cfghp97} 
and of Polyakov loop correlators~\cite{chp04}, we found a
very good agreement for large values of $R$ and a sizeable disagreement as $R$ decreases (see
fig.1a). This deviation (due to the peculiar geometry of the interface observable) 
cannot be due to boundary
corrections and is thus an intrinsic feature of the effective string description. 
In the case of square interfaces (i.e. $L_1=L_2\equiv R$) the NG effective string picture 
breaks down for values of $R$ smaller than $R\sim 2.5/\sqrt{\sigma}$ (see
fig.1a). It is important to notice that for square interfaces we see essentially no difference
between the prediction obtained using eq.(\ref{boskkp}) (black line in fig.1a) 
and the two loop approximation (red line in fig.1a).

\item
In order to disentangle between the two predictions one must study asymmetric interfaces because
higher orders in eq.(\ref{boskkp}) become more and more relevant as $u=L_2/L_1$ increases.
This is evident in fig.1b where the two curves are well separated. Looking at fig.1b we see that the
MC data show a much better agreement with the two loop approximation than with eq.(\ref{boskkp}).
This deviation from the behaviour of eq.(\ref{boskkp} may simply denote a breaking of the effective string
picture in this asymmetric limit. However it is also possible (and the persistent agreement with the two loop
approximation could  support this interpretation) that the effective string picture still holds 
but the corrections due to the anomaly (or equivalently, in the covariant
gauge framework, to the Liouville field) which affect the calculation of 
eq.(\ref{boskkp}) cannot be neglected.  
\item
Last, but not least, thanks to the improvement in the estimate of the interface string tension
$\sigma$ we could obtain new predictions 
for a few amplitude ratios which are 
more precise than any other existing estimate in the literature:

$$ \frac{T_c}{\sqrt{\sigma}}=1.235(2)$$
$$ \frac{m_{0++}}{\sqrt{\sigma}}=3.037(16)$$
$$ R_-\equiv \sigma \xi_{2nd}^2 =0.1024(5)$$

\end{itemize}

\section{Universal behaviour of interfaces in 2d and dimensional reduction of
  Nambu-Goto string}
It is interesting to study the $u\to \infty$ (i.e. $L_1<<L_2$)
limit of the eq.(\ref{boskkp}) for 
$d=3$. This corresponds to perform a dimensional reduction from $d=3$ to $d=2$
and allows us to obtain an effective description for bidimensional interfaces. 
The result is~\cite{bcf07}:
\begin{equation}
\label{2d}
Z=  \left(\frac{2}{\pi}\right)^ {\frac{1}{2}}\, L_3
\, m_{\mathrm{eff}} K_{1}\left(m_{\mathrm{eff}} L_2\right)~,
\hskip 1cm
m_{\mathrm{eff}} = \sigma L_1 \sqrt{1 - \frac{\pi}{3\sigma L_1^{2}}}
\end{equation}
Eq.(\ref{2d}) is most probably an exact result for any 2d spin model and not just an effective approximation. 
Indeed it agrees with an
exact calculation in the 2d Ising model. Moreover it arises from a simple universal model for 2d interfaces,
based on the first order treatment of an action given by the length of the interface.


  {\bf Acknowledgements}: M.P. acknowledges support from the Alexander von
Humboldt Foundation.

\end{document}